\documentclass[preprint,showpacs,preprintnumbers,amsmath,amssymb,endfloats*]{revtex4}
\usepackage{graphicx}

\newcommand{\bes}{\begin{eqnarray}}
\newcommand{\ees}{\end{eqnarray}}

\begin{document}

\thispagestyle{empty}
\title{
Theory confronts experiment in the Casimir force measurements:
quantification of errors and precision
}
\author{F.~Chen,${}^{1}$ 
   G.~L.~Klimchitskaya,${}^{2}$\footnote {On leave from
North-West Technical University, St.\ Petersburg, Russia}
U.~Mohideen,${}^{1}$\footnote{E-mail: umar.mohideen@ucr.edu}
 and
V.~M.~Mostepanenko${}^{2}$\footnote{On leave from Noncommercial
Partnership ``Scientific Instruments'', Moscow, Russia}
}

\affiliation{${}^{1}$Department of Physics, University of California,
Riverside, California 92521\\
${}^{2}$Departamento de F\'{\i}sica, Universidade Federal da Para\'{\i}ba,
C.P.5008, CEP 58059--970, Jo\~{a}o Pessoa, Pb-Brazil 
}

\begin{abstract}
We compare theory and experiment in the Casimir force measurement
between gold surfaces performed with the atomic force microscope.
Both random and systematic experimental errors are found leading
to a total absolute error equal to 8.5\,pN at 95\% confidence.
In terms of the relative errors, experimental precision of 1.75\% is 
obtained at the shortest separation of 62\,nm at 95\% confidence
level (at 60\% confidence the experimental precision of 1\% is
confirmed at the shortest separation). An independent
determination of the
accuracy of the theoretical calculations of the Casimir force and its
application to the experimental configuration is carefully
made. Special attention is paid to the sample-dependent
variations of the optical tabulated data due to the presence of grains,
contribution of surface plasmons, and errors introduced by the use of
the proximity force theorem. Nonmultiplicative and diffraction-type
contributions to the surface roughness corrections are examined.
The electric forces due to patch potentials resulting from the
polycrystalline nature of the gold films are estimated. The finite
size and thermal effects are found to be negligible. The theoretical
accuracy of about 1.69\% and 1.1\% are found at a separation 62\,nm
and 200\,nm, respectively. Within the limits of experimental and
theoretical errors very good agreement between experiment and theory 
is confirmed characterized by the root mean square deviation of about
3.5\,pN within all measurement range. The conclusion is made
that the Casimir force is stable relative to variations of the
sample-dependent optical and electric properties, which opens new
opportunities to use the Casimir effect for diagnostic purposes.
\end{abstract}


\maketitle


\section{Introduction}

In the last few years the Casimir effect \cite{1}, which is
a rare macroscopic manifestation of the boundary dependence of
the quantum vacuum, has attracted much experimental and theoretical attention
(see monographs \cite{2,3,4} and reviews \cite{5,6}). The spectrum
of the electromagnetic zero-point oscillations depends on the presence
of material bodies. In particular, the tangential component of the
electric field vanishes on the surfaces of two parallel plates
made of ideal metal (it is small if real metals are used). 
This leads
to changes in the zero-point oscillation spectrum compared to 
the case of free unbounded space and results in the 
attractive Casimir force acting normal to the surfaces
of the plates.

The Casimir effect finds many applications in quantum field
theory, condensed matter physics, elementary particle physics, gravitation
and cosmology \cite{2,3,4,5,6}. Recently many 
measurements of the Casimir force have been  performed 
\cite{7,8,9,10,11,12,13,14,15,16}. Their results have already 
been applied
in nanotechnology for the actuation of the novel microelectromechanical
devices, based entirely on the modification of the properties of
quantum vacuum \cite{17}, and for constraining predictions of
extra-dimensional physics with low compactification scales
\cite{14,16,18,19,20,21,22}.

Most theoretical papers on the Casimir effect deal with idealized
boundary conditions and perfectly shaped test bodies. 
Over the last 4 decades only a few
have considered  the corrections to the Casimir force such as due
to the finite conductivity of the boundary metal \cite{23,24,25},
distortions of the surface shape \cite{26,27} and nonzero temperature
\cite{28,29}. 
Comparison of the theory with the
results of modern Casimir force measurements demands careful treatment
of all these corrections.
Both the individual corrections and their combined effect
has to be evaluated (see Ref.~\cite{6} for review). 

The quantification of errors and precision in the measurements and
theoretical computations of the Casimir force is crucial for 
using the Casimir effect as a new test for extra-dimensional
physics and other extensions to the Standard Model. Nevertheless, there
is no general agreement on the achieved levels of experimental
precision and the extent of agreement between theory and experiment.
In the literature a variety of measures to 
characterize the experimental precision is used
and the extent of agreement between measurements and theory
ranges from 1\% \cite{8,10,11,16} to 15\% \cite{13} depending on
the measurement scheme and configuration. Very often, the confidence
levels and numerous background effects which may contribute to the
theoretical results are not considered.

In the present paper we perform a reanalysis of the experimental data
on the Casimir force measurements between Au surfaces \cite{11} and make 
a comparison with theory. In doing so we carefully calculate 
the original experimental precision without relation to the theory,
including the random absolute error at a 95\% confidence level, and the
absolute systematic error. The total absolute error of these Casimir
force measurements in the experiment of Ref.~\cite{11} is found to be
equal to $\Delta^{tot}\approx 8.6\,$pN at 95\% confidence. This corresponds
to approximately 1.75\% precision at the closest separation
$a\approx 62\,$nm (the 1\% precision at the closest separation indicated
in Ref~\cite{11} is obtained at 60\% confidence).
As a second step, the accuracy of the theoretical computations of the
Casimir force for the experimental configuration  \cite{11} is determined.
Special attention is paid to the possible sample-dependent variations
of the optical tabulated data due to the presence of grains, contribution
of the surface plasmons, and errors introduced by the use of the
proximity force theorem. The influence of the surface roughness is
carefully investigated including the nonmultiplicative contributions
and recently discussed diffraction-type effects \cite{30,31}. The
contribution of electric forces due to patch potentials resulting 
from the polycrystalline nature of the Au film is calculated
for the experimental configuration  \cite{11} at different separations.
The finite size and thermal effects are also considered and found
negligible in the experimental configuration of Ref.~\cite{11}.
The conclusion reached is that at the present state of our knowledge
the accuracies of theoretical computations
in application to the experimental configuration of Ref.~\cite{11}
are achievable
on the level of 1.69\% at a separation $z=62\,$nm and
1.1\% at a separation $z=200\,$nm.

The paper is organized as follows. In Sec.~II the experimental precision
of the Casimir force measurements at different confidence levels is
determined. Sec.~III is devoted to the computations of the Casimir force
with account of finite conductivity and grain structure of the metal
layers. The role of roughness including the nonmultiplicative
and diffraction-type effects is studied
in Sec.~IV. In Sec.~V both traditional and alternative thermal
corrections are discussed. Also the possible role of the electric forces
due to the patch potentials and finite size effects
are estimated. Sec.~VI contains the final numbers on theoretical 
accuracy and the comparison
of theory with experiment in the Casimir force measurement between two 
gold surfaces by means of an atomic force microscope \cite{11}. In Sec.~VII 
the final conclusions and some discussion are provided.

\section{Experimental precision in the Casimir force measurements
between two gold surfaces}

In Ref.~\cite{11} precision measurements of the Casimir force
between gold coated bodies, a plane plate and a sphere, were performed
using an atomic force microscope. The Casimir force was measured 
by averaging
30 scans over a surface separation region between 62--350\,nm
with 2583 points each (see Ref.~\cite{11} for all the details
of the measurement procedure). In the analysis below we neglect 
data from 3 scans due to excessive noise and use the 
data from the rest $n=27$ scans to find the quantitative characteristics of
the experimental precision in the Casimir force measurements at different
confidence levels.

We start with the random error and calculate the mean values of the
measured force at different separations $z_i$ within the region from
62\,nm to 350\,nm
\begin{equation}
\bar{F}^{exp}(z_i)=\frac{1}{n}
\sum\limits_{k=1}^{n}F_k^{exp}(z_i).
\label{eq1}
\end{equation}

An estimate for the variance of this mean is determined by \cite{32}
\begin{equation}
s_{\bar{F}}^2(z_i)=\frac{1}{n(n-1)}
\sum\limits_{k=1}^{n}\left[F_k^{exp}(z_i)-\bar{F}^{exp}(z_i)\right]^2.
\label{eq2}
\end{equation}
\noindent
Calculations using the measurement data $\{F_k^{exp}(z_i)\}$
show that $s_{\bar{F}}(z_i)$ do not depend sensitively on $z_i$.
The largest value $s_{\bar{F}}=2.8\,$pN is taken below as an estimate
for the variance of the mean force within the whole measurement
range.

According to Student's test for the truth of a hypothesis
\cite{32}, if the inequality
\begin{equation}
\frac{|\bar{F}^{exp}(z)-F(z)|}{s_{\bar{F}}}>t_{\alpha}^{\prime}
\equiv t_{1-\frac{1}{2}\alpha}
\label{eq3}
\end{equation}
\noindent
is fulfilled, the hypothesis that $F(z)$ is the true value of the
Casimir force at a separation $z$ must be rejected at a given
confidence level $\alpha$ (this is a two-tailed test as the
deviations $F(z)$ from $\bar{F}^{exp}(z)$ in two directions are
possible). Equivalently,  if the inequality
\begin{equation}
\frac{|\bar{F}^{exp}(z)-F(z)|}{s_{\bar{F}}}\leq 
t_{1-\frac{1}{2}\alpha}
\label{eq4}
\end{equation}
\noindent
is fulfilled, the hypothesis that $F(z)$ is the true value of the
Casimir force should be accepted at a confidence level $\beta=1-\alpha$. 

Usually in the tables for Student's $t$-distribution (see, e.g.,
Refs.~\cite{32,33}) the values of $t_p\equiv t_p(f)$ are presented,
where $p=1-\alpha/2=(1+\beta)/2$, $f=n-1$ is the number of degrees of
freedom, and $n$ is the number of measurements ($n=27$ in our case). 
Choosing $\beta=0.95$ (hypothesis is true at 95\% confidence)
we obtain $p=0.975$ and find from tables $t_p(f)=2.056$ \cite{33}.
Then from Eq.~(\ref{eq4}) it follows
\begin{equation}
|{\bar{F}}^{exp}(z)-F(z)|\leq\Delta^{rand}F^{exp}
\equiv s_{\bar{F}}t_p(f)\approx 5.8\,\mbox{pN},
\label{eq5}
\end{equation}
\noindent
where $\Delta^{rand}F^{exp}$ is the random absolute error of the
Casimir force measurements.
If we consider $\beta=0.6$ (hypothesis is true at 60\% confidence),
then $p=0.8$ and $t_p(f)=0.856$. In this case the random absolute error
of the Casimir force measurements is 
$\Delta^{rand}F^{exp}=2.8\times 0.856\,$pN$\approx 2.4\,$pN.
Note that if one would like to have $t_p(f)=1$ or
$t_p(f)=2$ (i.e. deviations of the true force value on either 
side of the mean not greater than one or two $s_{\bar{F}}$),
the confidence levels of $\beta=0.66$ or $\beta=0.94$, respectively,
should be chosen for the number of measurements $n=27$.

Now let us consider the systematic error. 
The main contributions to the systematic
error in the experiment of Ref.~\cite{11} are given by the error
in force calibration $\Delta_1^{syst}F^{exp}\approx 1.7\,$pN,
by the noise when the calibration voltage is applied to the cantilever
$\Delta_2^{syst}F^{exp}\approx 0.55\,$pN, by the instrumental
sensitivity $\Delta_3^{syst}F^{exp}\approx 0.31\,$pN, and by the restrictions
on computer resolution of data $\Delta_4^{syst}F^{exp}\approx 0.12\,$pN.
The maximal value of the systematic error is given by
\begin{equation}
\Delta^{syst}F^{exp}=\sum\limits_{i=1}^{4}
\Delta_i^{syst}F^{exp}\approx 2.7\,\mbox{pN}.
\label{eq6}
\end{equation}
\noindent
Finally, the maximum total absolute error of the Casimir force
measurements in the experiment of Ref.~\cite{11} is equal to
\begin{equation}
\Delta F^{exp}=\Delta^{rand}F^{exp}+
\Delta^{syst}F^{exp}\approx 8.5\,\mbox{pN}
\label{eq7}
\end{equation}
\noindent
at 95\% confidence (to be conservative, the errors are added
linearly rather than quadratically). 
At 60\% confidence the total absolute error of
the Casimir force measurements is $\Delta F^{exp}\approx 5.1\,$pN.
These absolute errors with their confidence levels are valid within the
whole measurement range from 62\,nm to 350\,nm. From Eq.~(\ref{eq5})
it follows that the true value of the Casimir force belongs to the
confidence interval
\begin{equation}
{\bar{F}}^{exp}(z)-\Delta F^{exp}\leq F(z)\leq
{\bar{F}}^{exp}(z)+\Delta F^{exp}
\label{eq8}
\end{equation}
\noindent
with a chosen confidence probability.

Another important characteristic of the experimental precision is the
relative error of the Casimir force measurements
$\delta F^{exp}(z)=\Delta F^{exp}/{\bar{F}}^{exp}(z)$ 
which is evidently separation-dependent.
At the shortest separation $z=62\,$nm the value of the mean force is
${\bar{F}}^{exp}=485.8\,$pN which leads to a relative error
of $\delta F^{exp}(z)\approx 1.75$\% computed at 95\% confidence.
If we restrict ourselves with a  60\% confidence, the relative error
of the Casimir force measurements at the shortest separation
 $\delta F^{exp}(z)=5.1/485.8\approx 1$\% is obtained as was
indicated in Ref.~\cite{11} without the detailed analysis of the
confidence levels. If we choose 95\% confidence, the relative errors
of the Casimir force measurements at separations 70\,nm, 100\,nm,
and 200\,nm are, respectively, 2.46\%, 5.9\%, and 37.3\%.
At 60\% confidence the relative errors of the Casimir force
measurements at the same separations are 1.47\%, 3.5\%, and 22.4\%,
respectively.

\section{Calculation of the Casimir force including the finite
conductivity and grain structure of gold layers}

For the  configuration of a large sphere of a radius $R$ above a plate
the Casimir force can be obtained by means of the Lifshitz formula,
derived originally for two parallel plates \cite{34}, along with 
use of the proximity force theorem \cite{35}
\begin{eqnarray}
&&
F_c(z)=\frac{\hbar R}{2\pi}
\int_{0}^{\infty}k_{\bot}dk_{\bot}
\int_{0}^{\infty}d\xi\left\{\ln\left[1-r_{\|}^2(\xi,k_{\bot})e^{-2zq}
\right]\right.
\nonumber \\
&&\phantom{aaaaaaaaaaa}
\left.
+\ln\left[1-r_{\bot}^2(\xi,k_{\bot})e^{-2zq}\right]\right\}.
\label{eq9}
\end{eqnarray}
\noindent
Here the reflection coefficients for two independent polarizations
are given by
\begin{eqnarray}
&&
r_{\|}^2(\xi,k_{\bot})=
\left[\frac{\varepsilon(i\xi)q-k}{\varepsilon(i\xi)q+k}\right]^2,
\qquad
r_{\bot}^2(\xi,k_{\bot})=\left(\frac{q-k}{q+k}\right)^2,
\nonumber \\
&&
q^2\equiv k_{\bot}^2+\frac{\xi^2}{c^2}, \qquad
k^2\equiv k_{\bot}^2+\varepsilon(i\xi)\frac{\xi^2}{c^2},
\label{eq10}
\end{eqnarray}
\noindent
$\varepsilon(\omega)$ is the dielectric permittivity of the 
gold layers on the sphere and the plate, and $z$ is the closest separation
distance between them. The thickness of gold coatings, used in
Ref.~\cite{11} (86.6\,nm), is much greater than the skin 
depth of the electromagnetic oscillations for all 
frequencies which make a significant contribution to the 
computation of the Casimir force. This allows one to use the 
properties of the bulk gold in all computations of the Casimir force.

The accuracy of Eq.~(\ref{eq9}) is restricted by the accuracy of the
proximity force theorem, which is, however, very high for the
experimental parameters of Ref.~\cite{11}. The error, introduced by
the proximity force theorem, is less than $z/R$ \cite{36,37}.
Taking into account the large value of sphere radius $R=95.65\,\mu$m,
used in Ref.~\cite{11}, the upper limit of this error is 0.06\%
at the shortest separation $z=62\,$nm and 0.2\% at separation
$z=200\,$nm (note that in Ref.~\cite{36} the Casimir force for 
the configuration of a sphere above a plate was 
precisely computed on the basis of the
first physical principles which makes it quite reliable as
a test of the proximity force theorem).

In Refs.~\cite{38,39} the computations of the Casimir force were
performed using Eqs.~(\ref{eq9}), (\ref{eq10}) and optical tabulated
data for gold \cite{40} (note that the transition coefficient from
energies to frequencies is given by 1\,eV$=1.52\times 10^{15}\,$rad/s).
The imaginary part of the dielectric permittivity, obtained using the
complex refractive index from the Tables \cite{40}, was used to compute
the dielectric permittivity along the imaginary frequency axis by
means of the dispersion relation. At $\omega<1.9\times 10^{14}\,$rad/s,
where the tabulated data are not immediately avaliable, they were
usually obtained (see, e.g., \cite{38,39})
by the extension from the region of higher frequencies by
means of the Drude dielectric function
\begin{equation}
\varepsilon(\omega)=1-\frac{\omega_p^2}{\omega(\omega+i\gamma)}\, ,
\label{eq11}
\end{equation}
\noindent
where the plasma frequency for Au is $\omega_p=1.37\times 10^{16}\,$rad/s, 
and $\gamma=5.32\times 10^{13}\,$rad/s is the relaxation parameter
describing the non-elastic electron-phonon interaction (note that
in the frequency region under consideration $\gamma\ll\omega$).
This procedure was used to
calculate the Casimir force including the effect of 
finite conductivity corrections 
of gold (see a few examples of the calculations in Sec.~IV,
Table~II and comparison between experiment and theory
in Sec.~VI). Later in this section we discuss the influence of possible
sample to sample variations of the optical tabulated data on the
values of the Casimir force and the applicability region of Eqs.~(\ref{eq9}),
(\ref{eq10}) involving the dielectric permittivity depending only on
frequency.

First, we would like to note that in the separation region
$200\,\mbox{nm}<z<350\,$nm the computational results obtained by
Eq.~(\ref{eq9}) combined with the optical tabulated data, are almost
exactly those obtained by the substitution into Eq.~(\ref{eq9})
of the plasma dielectric function for the metal
\begin{equation}
\varepsilon(\omega)=1-\frac{\omega_p^2}{\omega^2}\, .
\label{eq12}
\end{equation}
\noindent
In fact, both computations lead to results differing by less than
0.5\% within the mentioned separation region. What this means is that the
real part of $\varepsilon$ depending on only $\omega_p$ determines the
total value of the Casimir force in this region. The value of 
$\omega_p=2\sqrt{\pi N}e/\sqrt{m^{\ast}}$, where $N$ is the density of
conduction electrons, $m^{\ast}$ is their effective mass, is determined
by the preperties of the elementary cell. It cannot be influenced by
properties of sample such as the crystallite grain size 
or the presence of a small
concentration of impurities. This is the reason why the sample to sample
variations of the optical tabulated data cannot influence the value
of the Casimir force (\ref{eq9}) at separations $z\geq 200\,$nm. 

In the separation region $62\,\mbox{nm}<z<200\,$nm there are significant
deviations depending on whether 
the Casimir force (\ref{eq9}) is calculated using the 
optical tabulated data or  by use of the plasma dielectric
function (\ref{eq12}). In fact, in this separation region the small
imaginary part of $\varepsilon$ is influential and should be taken into
account. There is enough tabulated data in the 
optical Tables to compute the
Casimir force, so that it is not necessary to use any extension of data.
Note that the characteristic frequency corresponding to the largest
separation $z=200\,$nm is $\omega_c=c/(2z)=7.5\times 10^{14}\,$rad/s
(i.e. tabulated data for frequencies several times smaller are available).
At the same time, the characteristic frequency corresponding to the
shortest separation is 
$\omega_c=2.42\times 10^{15}\,\mbox{rad/s}\ll\omega_p$, so that the
region under consideration belongs to that of infrared optics \cite{41}.
Within the region (62--200)nm one may expect some small dependence
of the optical tabulated data on the size of the grain, 
presence of impurities etc. 
If this is indeed the case, the use of the tabulated data,
which are not relevant to the particular samples used in experiment,
might lead to the errors in computation of the Casimir force (\ref{eq9}).

To investigate this possibility, we consider the pure imaginary part
of the dielectric permittivity in the region of the infrared optics given
by \cite{42}
\begin{equation}
\mbox{Im}\varepsilon=\frac{\omega_p^2\nu}{\omega^3},
\label{eq13}
\end{equation}
\noindent
where $\nu$ is the relaxation parameter at high frequencies in the
region of infrared optics (note that it does not coincide with the
relaxation parameter $\gamma$ of the Drude model (\ref{eq11})
which describes the volume relaxation in the region of the normal skin effect).
The value of $\nu$ is determined by the processes of elastic scattering
of the electrons on impurities, on the boundary surfaces of the metal
and of the individual grains, and on other electrons \cite{42,43}.
The scattering of electrons on phonons also contributes 
to the value of $\nu$.
However, the frequency of the electromagnetic field is so high
that $\hbar\omega\gg k_BT_D$, where $T_D$ is the Debye temperature, so
the frequency of the electron-phonon collisions is the same as it is
at $T=T_D$ \cite{42}. 
It is important to note that of all the above processes, only the
contribution of the electron-electron collisions to $\nu$ is
frequency-dependent (and increases as $\omega^2$).

The main sample to sample dependence of the parameter $\nu$ is
determined by the sizes of grains and 
the density of impurities. To calculate this
dependence we use the following formula for the relaxation parameter
in the region of infrared optics \cite{42,44}
\begin{equation}
\nu=\omega_p\left(c_1+c_2\frac{\omega^2}{\omega_p^2}\right).
\label{eq14}
\end{equation}
\noindent
This formula leads to an approximate representation of the dielectric
permittivity of Au along the imaginary frequency axis given by
\begin{equation}
\varepsilon(i\xi)=1+\frac{\omega_p^2}{\xi^2}-
\frac{\omega_p^3}{\xi^3}
\left(c_1-c_2\frac{\xi^2}{\omega_p^2}\right),
\label{eq15}
\end{equation}
\noindent
where $c_1=0.0039$, $c_2=1.5$.
It is easily seen, that the substitution of Eq.~(\ref{eq15}) into
Eq.~(\ref{eq9}) leads approximately to the same result as the use of
the optical tabulated data. The errors due to use of Eq.~(\ref{eq15}) in
Eq.~(\ref{eq9}) instead of the optical tabulated data at separations
62\,nm, 70\,nm, 100\,nm, and 150\,nm,  are 0.45\%, 0.23\%, 0.09\%,
and 0.04\%, respectively.

Eq.~(\ref{eq15}) gives the possibility to estimate the influence of
the sizes of grains in the polycrystalline metal film in the experiment
of Ref.~\cite{11} on the value of the Casimir force (\ref{eq9}).
For this purpose, the experimental data of Ref.~\cite{45} are used
where the reflectance $R$ of Au films is measured as a function of
 the characteristic sizes of the grains. 

The analysis of the atomic force microscopy images (like the one in
Fig.~1 but on $1\times 1\,\mu\mbox{m}^2$ area) shows that the mean
size of grains in Ref.~\cite{11} is about 90\,nm (the sizes of the
typical grains are 77\,nm, 103\,nm, 94\,nm, 68\,nm, 88\,nm, 121\,nm
etc). According to Ref.~\cite{45}, the largest deviations of the
reflectance from the one given by the tabulated data \cite{40}, takes
place at shorter wavelengths. The shortest separation of $z=62\,$nm
in the experiment \cite{11} corresponds to the characteristic wavelength
$\lambda_c=2\pi c/\omega_c=4\pi z\approx 780\,$nm. For the films
containing grains of 45\,nm size (the largest ones studied in 
Ref.~\cite{45}) the reflectance at $\lambda\sim(750-800)$nm
is 0.8\% less than the one calculated from the tabulated data (notice that 
for smaller grains the difference of the reflectance obtained from the 
tabulated data is greater). Taking into account that the reflectance in the
region of the infrared optics is given by \cite{46}
\begin{equation}
R=1-4\mbox{Re}\frac{1}{\sqrt{\varepsilon}}=\frac{\nu}{\omega_p},
\label{eq16}
\end{equation}
\noindent
we find that the new value for the coefficient $c_1$ in Eqs.~(\ref{eq14}),
(\ref{eq15}) due to grains of 45\,nm size is ${\tilde{c}}_1=0.0059$.
Substituting the approximate Eq.~(\ref{eq15}) (with ${\tilde{c}}_1$
instead of $c_1$) into Eq.~(\ref{eq9}), one finds the values of the
correction factor to the Casimir force ${\tilde{\eta}}_c^A=0.439$
at $z=62\,$nm and ${\tilde{\eta}}_c^A=0.465$
at $z=70\,$nm. Comparing this with the results of the same approximate
computations using $c_1$ ($\eta_c^A=0.441$, respectively,
$\eta_c^A=0.467$), one can conclude that the grains of 45\,nm size
lead to less than 0.5\% decrease of the Casimir force magnitude.
Note that this is in fact the upper bound
for the influence of crystallite  grain size  
on the Casimir force in the experiment
of Ref.~\cite{11}, as the actual sizes of grains in \cite{11} were two 
times greater than 45\,nm.

The above calculations of the Casimir force including the effect 
of the real properties of Au films were performed on the basis of the Lifshitz
formula (\ref{eq9}), which does not take into account the effects of
spatial nonlocality (wavevector dependence of the dielectric permittivity). 
These effects may influence the Casimir force value in 
the region of the anomalous skin effect which is important 
for large
separations $z>2.36\,\mu$m \cite{47}, a region not relevant to the experiment
of Ref.~\cite{11}. Another separation region, where nonlocality
may lead to important contributions to the van der Waals force, is
$z<\lambda_p/(4\pi)\approx 10.9\,$nm  ($\lambda_p$
is the plasma wavelength) which corresponds to 
$\omega_c>\omega_p$ \cite{48}. Such high characteristic frequencies
lead to the propagation of surface plasmons. 
The effect of the surface plasmons,
however, does not contribute in the experiment of Ref.~\cite{11}
as the largest characteristic frequency there, calculated at $z=62\,$nm,
is 5.7 times less than $\omega_p$ [notice that the frequency
region ($5\omega_c,10\omega_c$) contributes only 0.19\% of the 
Casimir force value at separation $z=62\,$nm]. The contribution of
the surface plasmon for Au of about 2\% at a separation
$z=\lambda_p=137\,$nm, obtained in recent Ref.~\cite{49}, 
is explained by the use in \cite{49} of the
spatially nonlocal dielectric permittivity in the frequency region
of infrared optics where it is in fact local \cite{41,42,46}
(we would like to
point out that at the separation $z=\lambda_p$ the characteristic
frequency of the Casimir effect is equal not to $\omega_p$, as one
might expect, but $\omega_p/(4\pi)$].
As a result, the surface plasmons do not give any contribution to the
Casimir force in the experimental configuration of Ref.~\cite{11}.

\section{Surface roughness correction to the 
Casimir force and its calculation using different approaches}

It is well known that surface roughness corrections 
may play an important role 
in Casimir force calculations at separations less than 1\,$\mu$m
\cite{6}. At the shortest separations, the roughness correction
contributes 20\% of the measured force in experiments of Refs.~\cite{8,15,16}.
In the experiment of Ref.~\cite{11}, however, the roughness amplitude was 
decreased and the roughness contribution was made less than 1\% of
the measured force even at shortest separations. To obtain this
conclusion the simple stochastic model for the surface roughness and
the multiplicative approach to take into account different corrections
were used. Here we obtain more exact results for the contribution
of surface roughness to the Casimir force taking into account both
nonmultiplicative and correlation effects.

The topography of the Au coatings on the plate and sphere was
investigated using an atomic force microscope. A typical 3-dimensional
image resulting from the surface scan of $15\,\mu\mbox{m}\times 15\,\mu$m
area is shown in Fig.~1. As seen in this figure, the roughness is mostly
represented by the stochastically distributed distortions with the
typical heights of about 2--4\,nm, and rare pointlike peaks with the
heights up to 16\,nm. In Table~I the fractions $v_i$ of the surface
area, shown in Fig.~1, with heights $h_i$ are presented ($i=1,2,\ldots ,17$).
These data allow one to determine the zero roughness level $H_0$ relative to 
which the mean value of the function, describing roughness, is zero
(note that separations between different bodies in the Casimir force
measurements are usually measured between the zero roughness levels
\cite{6}):
\begin{equation}
\sum\limits_{i=1}^{17}\left(H_0-h_i\right)v_i=0.
\label{k1}
\end{equation}
\noindent
Solving Eq.~(\ref{k1}), one obtains $H_0\approx 2.734\,$nm. If the roughness
is described by the regular (nonstochastic) functions $Af(x,y)$,
where $|f(x,y)|\leq 1$, for the roughness amplitude it follows
$A=h_i^{max}-H_0=13.266\,$nm.

In the framework of the additive approach the values of the Casimir force
including the effect of finite conductivity $F_c(z)$, obtained in Sec.~III,
Eq.~(\ref{eq9}), can be used to calculate the effect of roughness.
For this purpose, the values of $F_c$ should be geometrically
averaged over all different possible separations between the rough
surfaces weighted with the probability of each separation \cite{6,8,16}
\begin{equation}
F_{c,r}(z)=
\sum\limits_{i,j=1}^{17}v_iv_jF_c(z+2H_0-h_i-h_j).
\label{k2}
\end{equation}
\noindent
Note that Eq.~(\ref{k2}) is not reduced to a simple multiplication
of the correction factors due to finite conductivity and surface
roughness but takes into account their combined (nonmultiplicative)
effect.

An alternative method of calculating the corrections due to the 
stochastic surface roughness
was used in Ref.~\cite{11}. According to the results of Ref.~\cite{50},
the Casimir force between a plate and a sphere made of ideal metal and
covered by a stochastic roughness with an amplitude $A_{st}$ is given by
\begin{equation}
F_r(z)=F_0(z)\left[1+6\left(\frac{A_{st}}{z}\right)^2
+45\left(\frac{A_{st}}{z}\right)^4\right],
\label{k3}
\end{equation}
\noindent
where $F_0(z)=-\pi^3\hbar cR/(360z^3)$ is the Casimir force between
perfectly shaped plate and sphere of radius $R$.
Then the Casimir force including both the finite conductivity
of the boundary metal and surface roughness can be calculated as 
\begin{equation}
F_{c,r}^{m}(z)=F_c(z)\left[1+6\left(\frac{A_{st}}{z}\right)^2
+45\left(\frac{A_{st}}{z}\right)^4\right],
\label{k4}
\end{equation}
\noindent
i.e. by means of the multiplicative procedure.

The variance of the random process describing the stochastic roughness is
found by the formula
\begin{equation}
\delta_{st}^2=\sum\limits_{i=1}^{17}\left(H_0-h_i\right)^2 v_i.
\label{k5}
\end{equation}
\noindent
Using the data from Table~I, one obtains the values for variance
$\delta_{st}\approx 0.837\,$nm and for the amplitude of a random process
$A_{st}=\sqrt{2}\delta_{st}\approx 1.18\,$nm. This value is slightly 
larger than the one obtained in Ref.~\cite{11} on the basis of less
complete data on roughness topography.

Now we are in a position to compare the contribution of the surface
roughness computed by Eq.~(\ref{k2}), taking into account the
combined effect of the roughness and finite conductivity, and by the
multiplicative procedure of Eq.~(\ref{k4}). In Table~II the results for
the correction factors $\eta_c=F_c/F_0$, $\eta_r=F_r/F_0$,
 $\eta_{c,r}=F_{c,r}/F_0$, and $\eta_{c,r}^{m}=\eta_c\eta_r$ are
presented at the shortest separations $z=62\,$nm,\ 70\,nm,\ 80\,nm, and
90\,nm, where the roughness corrections play some role.
As is seen from Table~II, both approaches lead to practically
coincident results for the roughness correction factors due to the combined
effect of finite conductivity and surface roughness. This means that
for such small roughness as in Ref.~\cite{11} the multiplicative
procedure is quite satisfactory (for larger roughness amplitudes,
however, the nonmultiplicative contributions may be essential \cite{8,16}).
Note also that for $A_{st}\approx 1.18\,$nm the fourth order term in
Eq.~(\ref{k4}) practically does not contribute even at shortest separations 
and can be neglected as was done in Ref.~\cite{11}.

Both Eqs.~(\ref{k2}) and (\ref{k4}) used above are based on the
approximation of additive summation and do not take into account
the diffraction-type effects which arise in the case of roughness
described by the periodic functions with small periods $\lambda<z$
\cite{30} or by the stochastic functions with small correlation length
\cite{31}. To estimate the value of the correlation length in our case,
we consider the set of cross sections of the roughness image shown
in Fig.~1.

In Fig.~2 two typical cross sections are presented, one at fixed $x$
(a) and the other one at fixed $y$ (b). We have performed the Fourier
analysis of the functions, as in Figs.~2,a,b, along the lines of
Ref.~\cite{27}. It was found that the Fourier harmonics, giving the
major contribution to the result, are characterized by significantly
greater periods than the mean distance between the neighbouring peaks in
Figs.~2,a,b which is equal, approximately, to 180\,nm.

To obtain an estimate for the upper limit of the contribution of the
diffraction-type effects in the above roughness analysis, we use the
correlation length $l_{corr}=200\,$nm (slightly larger than the mean
distance between peaks) and consider the periodic function with this period 
(clearly, the diffraction-type effects are greater for a periodic
function with a period $l_{corr}$ than for the random function
with a correlation length $l_{corr}$). With this the diffraction-type effects
can be computed in the framework of the functional approach developed
in Ref.~\cite{30}. At a shortest separation $z=62\,$nm one obtains 
$z/l_{corr}\approx 0.31$. Then for the coefficient $c_{corr}$ in the
expression
\begin{equation}
\eta_r^{corr}=1+6c_{corr}\left(\frac{A_{st}}{z}\right)^2,
\label{k6}
\end{equation}
\noindent
taking the diffraction-type effects into account, 
from Fig.~2 of Ref.~\cite{30}
it follows $c_{corr}\approx 1.1$. As a result, using the upper
limit for the contribution of the diffraction effects one obtains
$\eta_r^{corr}\approx 1.0024$, i.e. only 0.02\% difference with the value
of $\eta_r$ in Table~II obtained by neglecting the diffraction
effects. At larger separations the diffraction effects lead to larger
contribution to the roughness corrections. For example, at a
separation $z=90\,$nm we have $z/l_{corr}\approx 0.45$, 
$c_{corr}\approx 1.28$, and $\eta_r^{corr}\approx 1.0013$, i.e. 0.03\%
difference with the result of Table~II. At larger separations, however,
the roughness correction itself is even more negligible than at the
shortest separations.

To conclude, the surface roughness contribution in the experiment of
Ref.~\cite{11} does not exceed 0.24\% of the Casimir force at the shortest
separation $z=62\,$nm. The diffraction-type effects, which were not taken
into account in Eqs.~(\ref{k2}), (\ref{k4}), are shown to contribute less
than one tenth of this result.

\section{Contributions of the thermal corrections, residual electric forces
and finite sizes of the plate}

Although the experiment of Ref.~\cite{11} was performed at room
temperature $T=300\,$K, all the above computations were done at zero
temperature. The thermal Casimir force $F_{c}(z,T)$ is given by Eq.~(\ref{eq9})
where integration in continuous $\xi$ is changed to a summation over
the discrete Matsubara frequencies $\xi_l=2\pi k_BTl/\hbar$ according to
\[
\int_{0}^{\infty}d\xi\to\frac{2\pi k_BT}{\hbar}\sum\limits_{l=0}^{\infty}
{\vphantom{\sum}}^{\prime},
\]
\noindent
leading to the Lifshitz formula for the thermal Casimir force.
Here prime refers to the addition of a multiple 1/2 near the term
with $l=0$. When $T\to 0$, $F_c(z,T)\to F_c(z,0)=F_c(z)$, where $F_c(z)$
is given by Eq.~(\ref{eq9}).

The magnitude of the relative thermal correction to the Casimir force
can be computed by the formula
\begin{equation}
\delta_TF_c(z,T)=\frac{F_c(z,T)-F_c(z)}{F_c(z)}
\equiv\frac{\Delta_TF_c(z,T)}{F_c(z)}.
\label{eq23}
\end{equation}

Recently, there has been extensive discussion in literature on the
correct calculation procedure for the thermal Casimir force $F_c(z,T)$
\cite{47}. In Refs. ~\cite{51,52} the dielectric permittivity of the
plasma model (\ref{eq12}) was substituted into the Lifshitz formula
for $F_c(z,T)$. This approach, which was later called ``traditional''
\cite{15}, leads to the thermal corrections $\Delta_T^{tr}F_c$,
$\delta_T^{tr}F_c$. It is consistent with thermodynamics and agrees
with the limiting case of the ideal metal.  In the region of infrared
optics the same results were obtained in the framework of the impedance
approach which does not consider the fluctuating electromagnetic field
inside the  metal and takes into account the realistic properties of the
metal by means of the Leontovich boundary condition \cite{47,53}.
Within the separation distances of Ref.~\cite{11}, the traditional 
thermal corrections are very small. As an example, at a separation
$z=100\,$nm and $T=300\,$K one has $\delta_T^{tr}F_c\approx 0.007$\%,
and $\delta_T^{tr}F_c\approx 0.03$\%, 0.1\% at  separations 
$z=200\,$nm, respectively, 300\,nm 
\cite{54} (to compare, in the case of ideal metals the same corrections,
found in the framework of the thermal quantum field theory, are equal
to 0.003\%, 0.024\%, and 0.08\%, 
respectively, i.e. the results for real metals
approach the results for ideal ones with the increase of separation
\cite{29,54}). Thus, the traditional thermal corrections are negligible
in the measurement range of experiment \cite{11}
(the contribution of the relaxation processes to the magnitude of
these corrections, which can be computed by taking into account the
small real part of the surface impedance, is much less than the 
corrections).

Alternatively, in Refs.~\cite{55,56} the dielectric permittivity of the
Drude model (\ref{eq11}) was used to calculate $F_c(z,T)$. In this approach,
there is no continuous transition between the cases of real and ideal
metal. At the high temperature limit, the Casimir force between real
metals was found equal to one half of the result obtained for the ideal
metal (independently of how high the conductivity of real metal is).
The thermal corrections, computed in the framework of the alternative 
approach \cite{55,56}, are quite different from those obtained from  
the traditional approach. 
To find the magnitude of these corrections \cite{54},
one should substitude into Eq.~(\ref{eq23})
\begin{equation}
\Delta_T F_c(z,T)\equiv \Delta_T^{(1)}F_c(z,T)\approx
\Delta_T^{tr}F_c(z,T)-\frac{k_BTR}{8a^2}
\int_{0}^{\infty}ydy\ln\left[1-r_{\bot}^2(0,y)e^{-y}\right],
\label{eq24}
\end{equation}
\noindent
where $r_{\bot}^2(0,y)$ is obtained by the substitution of Eq.~(\ref{eq12}) 
into Eq.~(\ref{eq10}). After calculations, one obtains that the
alternative relative thermal correction increases from
$\delta_T^{(1)}F_c\approx 1.1$\% and 1.3\% at separations $z=62\,$nm,
respectively, 70\,nm to $\delta_T^{(1)}F_c\approx 8$\% at a
separation $z=350\,$nm.

Another alternative thermal correction suggested in literature \cite{57}
is also based on the substitution of the Drude dielectric function
(\ref{eq11}) into the Lifshitz formula for $F_c(z,T)$ but with a
modified zero-frequency contribution for the perpendicular mode (in 
Ref.~\cite{57} this contribution is postulated to be of the same value as
for an ideal metal). The alternative thermal correction of Ref.~\cite{57}
is given by \cite{54}
\begin{equation}
\Delta_T F_c(z,T)\equiv \Delta_T^{(2)}F_c(z,T)\approx
\Delta_T^{(1)}F_c(z,T)+\frac{k_BTR}{8a^2}
\int_{0}^{\infty}ydy\ln\left(1-e^{-y}\right).
\label{eq25}
\end{equation}
\noindent
As a result, the relative alternative thermal correction of this kind
takes values $\delta_T^{(2)}F_c\approx (2.1-2.2)$\% at all separations
from $z=62\,$nm to $z=350\,$nm, i.e. slightly larger than the experimental
precision at the shortest separations. 

As was shown in Ref.~\cite{58}, both alternative thermal corrections
of Refs.~\cite{55,56} and of Ref.~\cite{57} are not consistent  with
thermodynamics leading to the violation of the Nernst heat theorem.
Recently they were found to be in disagreement with the
precision measurement of the Casimir force using a 
microelectromechanical torsional oscillator \cite{15}. In Sec.~VI we will 
discuss the influence of the alternative thermal corrections on the
comparison of theory and experiment in the Casimir force
measurement of Ref.~\cite{11}.

In the rest of this section we discuss the probable contribution of the
residual electric forces and the finite sizes of the plate on the
Casimir force. As was noted in Ref.~\cite{11}, the electrostatic force 
due to the residual potential difference between the plate and the sphere 
has been lowered to negligible levels of $\ll 1$\% of the Casimir force
at the closest separations. In recent Ref.~\cite{59} it was argued,
however, that the spatial variations of the surface potentials due to
the grains of polycrystalline metal (the so called ``patch
potentials'') may mimic the Casimir force. Here we apply the general
results of Ref.~\cite{59} to the experiment of Ref.~\cite{11} and
demonstrate that the patch effect does not make significant contributions.

According to Ref.~\cite{59}, for the configuration of a sphere above a plate
the electric force due to random variations in patch potentials is
given by
\begin{equation}
F_p(z)=-\frac{4\pi\varepsilon_0\sigma_v^2R}{k_{\max}^2-k_{\min}^2}
\int_{k_{\min}}^{k_{\max}}\frac{k^2e^{-kz}dk}{\sinh kz},
\label{eq26}
\end{equation}
\noindent
where $\sigma_v$ is the variance of the potential distribution,
$k_{\max}$ ($k_{\min}$) are the magnitudes of the extremal
wavevectors corresponding to minimal (maximal) sizes of grains,
and $\varepsilon_0$ is the dielectric permittivity of free space.
The work functions of gold are $V_1=5.47\,$eV, $V_2=5.37\,$eV,
and $V_3=5.31\,$eV for different crystallographic surface orientations
(100), (110), and (111), respectively. Assuming 
equal areas of these crystallographic planes one obtains
\begin{equation}
\sigma_v=\frac{1}{\sqrt{2}}\left[
\sum\limits_{i=1}^{3}(V_i-\bar{V})^2\right]^{1/2}
\approx 80.8\,\mbox{mV}.
\label{eq27}
\end{equation}

Using the atomic force microscopy images discussed in Sec.~III, the
extremal sizes of grains in gold layers covering the test bodies
were determined $\lambda_{\min}\approx 68\,$nm, and
$\lambda_{\max}\approx 121\,$nm. This leads to 
$k_{\max}\approx 0.092\,\mbox{nm}^{-1}$ and
$k_{\min}\approx 0.052\,\mbox{nm}^{-1}$. Note that these grain
sizes are of the same order as the thickness of the film.
The computations by Eq.~(\ref{eq26}) using the above data
lead to the ``patch effect'' 
electric forces $F_p/R\approx -1.15\times 10^{-8}\,$N/m
and $ -1.25\times 10^{-10}\,$N/m at separations $z=62\,$nm and
$z=100\,$nm, respectively. Comparing the obtained results with
the values of the Casimir force at the same separations
($F_c/R\approx -5.06\times 10^{-6}\,$N/m, respectively,
$ -1.48\times 10^{-6}\,$N/m), we conclude that the electric force
due to the patch potentials contributes only 0.23\% and 0.008\%
of the Casimir force at separations $z=62\,$nm, respectively,
$z=100\,$nm (at a separation $z=200\,$nm the patch effect contributes
only $7\times 10^{-7}$\% of the Casimir force. So a rapid decrease
of the contribution of the electric force with an increase of a separation
is explained by the exponential decrease of the integral in Eq.~(\ref{eq26})
if one substitutes the physical values of the integration limits based
on the properties of gold films used in experiment of Ref.~\cite{11}.

A more important contribution of the patch electric forces may be expected 
in the scanning of a sharp tip of the atomic force microscope at a height 
of about 10--20\,nm above a plate. In this case the electric forces 
are comparable with the van der Waals forces complicating the
theoretical interpretation of force-distance relations \cite{60}.

Let us finally estimate the theoretical error caused by finiteness of
the plate used in the experiment of Ref.~\cite{11}. Eq.~(\ref{eq9})
was derived for the plate of infinite radius. In fact, the radius of
the plate used in the experiment \cite{11} is $L=5\times 10^{-3}\,$m.
In this case, the Casimir force can be obtained by the formula \cite{61}
\begin{equation}
F_c^{fin}(z)=F_c(z)\beta(z)=\left[1-\frac{z^3}{R^3}\left(1-
\frac{1}{\sqrt{1+L^2/R^2}}\right)^{-3}\right]F_c(z).
\label{eq28}
\end{equation}
\noindent
To calculate $\beta(z)$, we put $z=350\,$nm (to make this factor
maximally distinct from unity) and obtain
\[
\beta(z)\approx 1-8\frac{z^3R^3}{L^6}\approx 1-2.2\times 10^{-17},
\]
\noindent
i.e. the finiteness of the plate size is too small  
to give any meaningful  contributions to the Casimir force.

\section{Theoretical accuracy and comparison of theory and
experiment}

Now we are in a position to list all the sources of errors in the
theoretical computation of the Casimir force
$F_{c,r}$ given by Eq.~(\ref{k2}), to find the final
theoretical accuracy and to consider the comparison of theory
and experiment.

The main error, which arises from Eqs.~(\ref{eq9}), (\ref{k2})
 when one substitutes 
the experimental data, is due to the errors in determination of
separation $z$ and sphere radius $R$. In terms of dimensionless
variables, $F_{c,r}(z)$ is proportional to $R$ and inversely proportional
to $z^3$ which results in
\begin{equation}
\delta^{(1)}F_{c,r}=\frac{\Delta F_{c,r}}{F_{c,r}}\approx
\frac{\Delta R}{R}+3\frac{\Delta z}{z}.
\label{eq29}
\end{equation}
\noindent
In Ref.~\cite{11} the absolute separations were determined by means of 
the electric measurements which allowed the determination of  
the average 
separation distance on contact $z_0$ 
with the absolute error $\Delta^{el}z_0\approx 1\,$nm.
Contrary to the opinion expressed in Ref.~\cite{62},
this error should not be transferred  to all separations
leading to rather large contribution to $\delta^{(1)}F_{c,r}$ of about
$3\Delta^{el}z_0/z\approx 4.8$\% at the shortest separations.
Note that we are comparing not one experimental point with
one theoretical value, but the experimental force-distance relation
with the theoretical one computed on the basis of a fundamental theory.
Thus, an additional fit should be made, with $z_0$ as a
fitting parameter changing within the limits 
($z_0-\Delta^{el}z_0,z_0+\Delta^{el}z_0$) to achieve the smallest root
mean square (r.m.s.) deviations between experiment and theory
\begin{equation}
\sigma_M=\left\{\frac{1}{M}\sum\limits_{i=1}^{M}
\left[F_{c,r}(z_i)-{\bar{F}}^{exp}(z_i)\right]\right\}^{1/2},
\label{eq30}
\end{equation}
\noindent
where ${\bar{F}}^{exp}(z_i)$ was defined in Eq.~(\ref{eq1}), 
$F_{c,r}(z_i)$ were computed by Eq.~(\ref{k2}), and $M$ is the number
of experimental points under consideration. If, as usual, we consider
two hypotheses as equivalent when they lead to the r.m.s. deviations
differing for less than 10\%, this results in decrease of the
error in determination of absolute separations up to 
$\Delta z \approx 0.15\,$nm.

It is important to underline that the verification of the hypothesis is
performed within different separation intervals (i.e. the total number
$M=N=2583$ experimental points within the whole separation range
from 62\,nm to 350\,nm, $M=1270$ points belonging to the interval
62--210\,nm, and $M=600$ points at separations less than the plasma 
wavelength $\lambda_p=136\,$nm). The above value for $\Delta z$ is 
almost one and the same in all the separation intervals. The obtained
values of the r.m.s. deviations between theory and experiment are
$\sigma_N\approx 3.4\,$pN, $\sigma_{1270}\approx 3.2\,$pN, and
$\sigma_{600}\approx 3.8\,$pN. These values are rather homogeneous
demonstrating good agreement between theory and experiment independently
of the chosen separation region.

The radius of the sphere was measured more precisely than in Ref.~\cite{11}
with a result $2R=191.3\pm 0.3\,\mu$m. Using this together with
$\Delta z=0.15\,$nm, one obtains from Eq.~(\ref{eq29}) at the shortest
separations $\delta^{(1)}F_{c,r}\approx 0.88$\%.

Now let us list the other contributions to the theoretical error of the
Casimir force computations at the shortest separation $z=62\,$nm and
indicate their magnitude. According to the results of Sec.~III, the
sample to sample variations of the optical tabulated data may lead to the
decrease of the Casimir force magnitude for no more than
$\delta^{(2)}F_{c,r}\approx 0.5$\%. The use of  the proximity force
theorem at $z=62\,$nm leads to very small error of about 
$\delta^{(2)}F_{c,r}\approx 0.06$\% (Sec.~III). The corrections due to
the surface roughness are already incorporated in the theoretical
expression (\ref{k2}) but the diffraction-type effects may contribute
up to $\delta^{(4)}F_{c,r}\approx 0.02$\% (Sec.~IV). The effect of
electric forces due to the patch potentials contribute a maximum
$\delta^{(5)}F_{c,r}\approx 0.23$\% at the shortest separations,
as was shown in Sec.~V. The corrections due to the surface plasmons 
and finite size of the plate 
are negligible for the separation distances and experimental
configuration used in Ref.~\cite{11} (see Secs.~III and V).

Special attention should be paid to the thermal corrections to the
Casimir force. According to the results of Sec.~V, the contribution of
the traditional thermal correction at the shortest separation is
negligible. At larger separations it may be incorporated into the
theoretical expression for the force. As to the alternative thermal 
corrections of Refs.~\cite{55,56,57}, which contribute of about
(1--2)\% of the Casimir force at the separation $z=62\,$nm, they have
been already ruled out both experimentally and theoretically (see Sec.~V).
If we would include any of these corrections into the theoretical
expression for the Casimir force, this results in the increase of the
r.m.s. deviation between theory and experiment which cannot be
compensated by shifts of the separation distance in the limit of
error $\Delta^{el}z_0$. In view of the above, we exclude the contributions
from these hypothetical corrections from our error analysis.

The upper limit for the total theoretical error at a separation
$z=62\,$nm can be found by the summation of the above contributions
\begin{equation}
\delta F_{c,r}=\sum\limits_{i=1}^{5}\delta^{(i)}F_{c,r}
\approx 1.69\mbox{\%},
\label{eq31}
\end{equation}
\noindent
which is a bit more accurate than the total experimental relative error
at the shortest separation equal to 1.75\% at 95\% confidence
(see Sec.~II).

Note that with the increase of separation the experimental relative
error quickly increases to 37.3\% at a separation $z=200\,$nm.
At the same time, the theoretical error is slowly decreasing with 
increasing separation. Thus, at $z=200\,$nm the above contributions
to the theoretical error of the Casimir force computations take values
$\delta^{(1)}F_{c,r}\approx 0.38$\%, $\delta^{(2)}F_{c,r}\approx 0.5$\%,
$\delta^{(3)}F_{c,r}\approx 0.21$\%, $\delta^{(4)}F_{c,r}\approx 0.026$\%,
$\delta^{(5)}F_{c,r}\approx 0.000$\%.
As a result, the total theoretical error at $z=200\,$nm is 
$\delta F_{c,r}\approx 1.1$\%.

The obtained results demonstrate very good agreement between theory and 
experiment within the limits of both experimental and theoretical errors.

\section{Conclusions and discussion}

In this paper we have performed a detailed comparison of experiment
and theory in the Casimir force measurement between the gold coated
plate and sphere by means of an atomic force microscope \cite{11}.
The random error of the experimental values of the Casimir force
was found to be $\Delta^{rand}F^{exp}\approx 5.8\,$pN at 95\%
confidence (at 60\% confidence the value 
$\Delta^{rand}F^{exp}\approx 2.4\,$pN was obtained). Together with
the systematic error $\Delta^{syst}F^{exp}\approx 2.7\,$pN, this
leads to the total absolute error of the Casimir force measurements
in Ref.~\cite{11} $\Delta{}F^{exp}\approx 8.5\,$pN at 95\%
confidence. In terms of the relative errors, the experimental
precision at the shortest reported separation is equal to 1.75\% (1\%)
at 95\% (60\%) confidence level.

In order to find the theoretical accuracy of the Casimir force 
calculations in the experimental configuration
of Ref.~\cite{11}, many corrections to the ideal Casimir
force were analysed. The correction due to the finite conductivity 
of gold was computed by the use of the optical tabulated data of the
complex refractive index. The results were compared with those
computed by the use of the plasma dielectric function and
found to coincide for the surface separation range (200--350)nm.
At shorter separations the use of the optical tabulated data more 
accurately represents the dielectric properties.
A special model was presented, which allows to
take into account the sample to sample variations of the optical
tabulated data due to the sizes of grains and impurities.
It was shown that the error introduced by the grains of 45\,nm size
(even smaller than those in the experiment of Ref.\cite{11}) does
not exceed 0.5\% of the Casimir force. The influence of the surface
plasmon in the separation region of the experiment \cite{11} was
found to be negligible.

The surface roughness of the test bodies, used to measure the Casimir 
force, was carefully investigated by means of the atomic force
microscope with a sharp tipped cantilever instead of a
large sphere. The obtained profiles of roughness topography 
allowed calculation of the roughness corrections to the Casimir force in
the framework of both multiplicative and nonmultiplicative approaches.
The minor differences in the size of the effect are found only
at the shortest separation. The correlation length of the surface 
roughness on the test bodies was estimated and the diffraction-type
effects were computed. At the shortest separation the roughness
correction contributes 0.24\% of the Casimir force with account of
diffraction-type effects (and 0.22\% with no account of diffraction). 

The electric forces caused by the spatial variations of the surface 
potentials due to the size of grains were investigated for
the experimental configuration of Ref.~\cite{11}. They were
shown to contribute 0.23\% of the Casimir force at the shortest
separation, and this contribution quickly decreases
with an increase of separation. Several other effects (such as thermal
corrections, corrections due to the finiteness of the plate and
due to the deviation from the proximity force theorem) were
investigated and found to make only negligible contributions.

The final theoretical accuracy of the Casimir force calculations in the
experimental configuration of Ref.~\cite{11} is 1.69\% at the
shortest separation $z=62\,$nm and 1.1\% at a separation $z=200\,$nm.
In the limits of both experimental and theoretical errors, very good 
agreement between theory and experiment was demonstrated characterized
by the r.m.s. deviation of about 3.5\,pN (less than 1\% of the
measured force at a shortest separation) which is almost independent
of the separation region and the number of the experimental points.
The above analysis does not support the conclusion of Ref.~\cite{62}
that to achieve the experiments on the Casimir effect with 1\%
precision it is necessary to measure the separation on contact
$z_0$ with atomic precision.

The obtained results demonstrate that in fact the Casimir force is
more stable, than one might expect, to some delicate properties of
the metallized test bodies like the variations of the optical data,
patch potentials, correlation effects of roughness etc. These
properties may change from sample to sample leaving the basic
character, and even the values of the Casimir force within some definite
separation region, almost unchanged. The stability of the Casimir force 
opens new opportunities to use the Casimir effect 
as a test for long-range hypothetical interactions and for the 
diagnostic purposes. For example, some kind of the inverse problem 
could be utilized,
i.e. the measured force-distance relations be exploited to determine
the fundamental characteristics of solids (such as the plasma
frequency).

\section*{Acknowledgments}

This work was supported by the National Institute for Standards and
Technology and a University of California and
Los Alamos National Laboratory grant through the LANL-CARE 
program.
G.L.K. and V.M.M. were 
also partially supported by CNPq and Finep (Brazil).

\begin{table}
\caption{Fractions $v_i$ of the surface area covered
by roughness with heights $h_i$.}
\begin{ruledtabular}
\begin{tabular}{rcc}
i & $h_i$\,(nm) & $v_i$ \\
\hline
1 & 0 & $1.06\times 10^{-3}$ \\
2 & 1 & $5.086\times 10^{-2}$ \\
3 & 2 & $0.33511$ \\
4 & 3 & $0.45863$ \\
5 & 4 & $0.13695$ \\
6 & 5 & $1.586\times 10^{-2}$ \\
7 & 6 & $1.24\times 10^{-3}$ \\
8 & 7 & $1.6\times 10^{-4}$ \\
9 & 8 & $4\times 10^{-5}$ \\
10 & 9 & $2\times 10^{-5}$ \\
11 & 10 & $1\times 10^{-5}$ \\
12 & 11 & $1\times 10^{-5}$ \\
13 & 12 & $1\times 10^{-5}$ \\
14 & 13 & $1\times 10^{-5}$ \\
15 & 14 & $1\times 10^{-5}$ \\
16 & 15 & $1.2\times 10^{-5}$ \\
17 & 16 & $8\times 10^{-6}$ \\
\end{tabular}
\end{ruledtabular}
\end{table}
\begin{table}
\caption{Corrections factors to the ideal Casimir force
at different separations due to finite conductivity
($\eta_c$), surface roughness ($\eta_r$) and both
finite conductivity and surface roughness ($\eta_{c,r}$ 
and $\eta_{c,r}^{m}$ in the method of the geometrical
averaging and in the multiplicative approach, respectively).}
\begin{ruledtabular}
\begin{tabular}{lcccc}
& $z=62\,$nm &$z=70\,$nm &$z=80\,$nm &$z=90\,$nm \\
\hline
$\eta_c$ & 0.4430 &0.4681 & 0.4964 &0.5218 \\
$\eta_r$ & 1.0022 & 1.0017 & 1.0013 & 1.0010 \\
$\eta_{c,r}$ & 0.4436 & 0.4687 & 0.4669 & 0.5223 \\
$\eta_{c,r}^{m}$ & 0.4440 & 0.4689 & 0.4670 & 0.5223 \\
\end{tabular}
\end{ruledtabular}
\end{table}
\begin{figure*}
\vspace*{-3cm}
\includegraphics{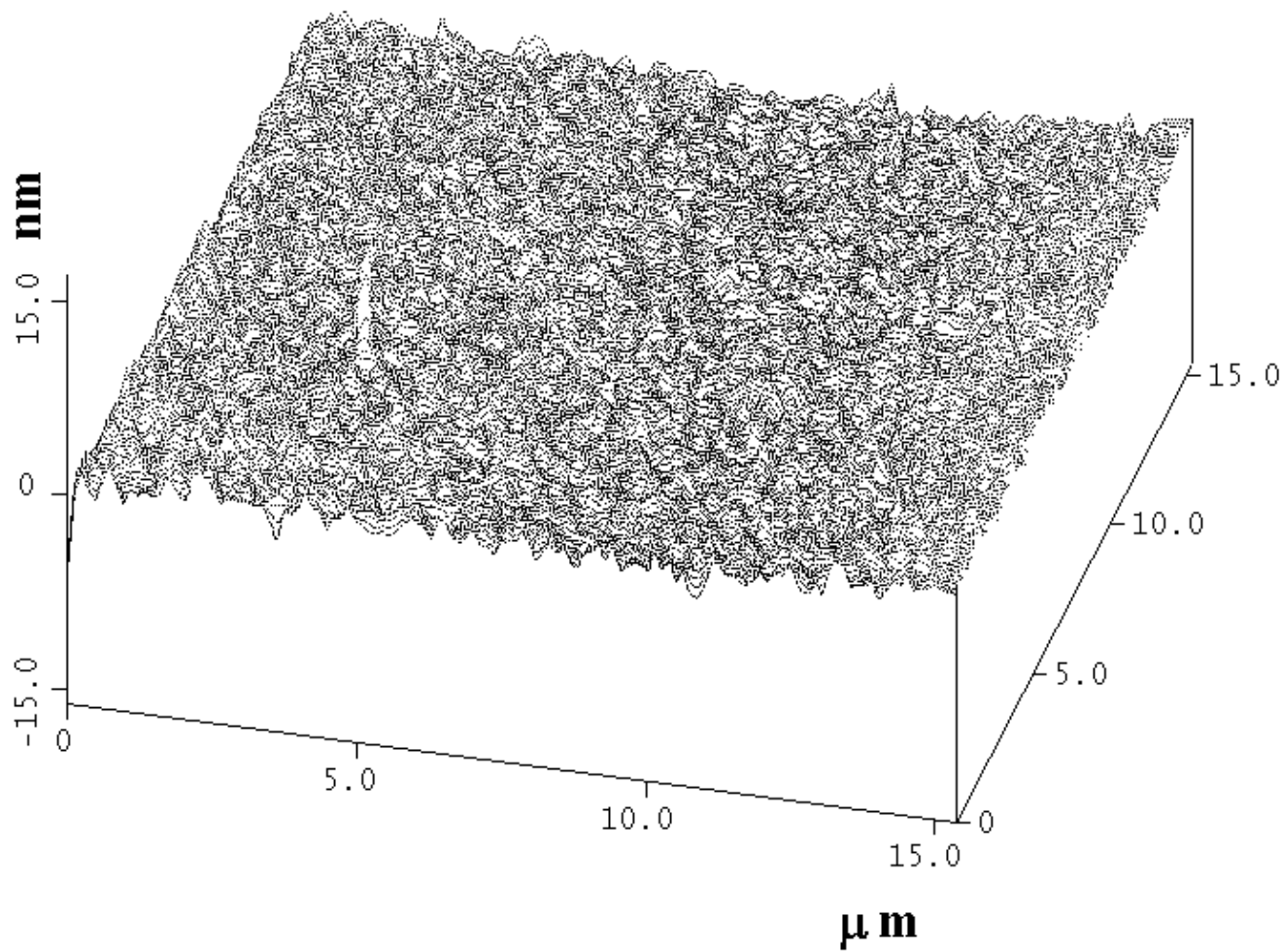}
\vspace*{-7cm}
\caption{$15\times 15\,\mu\mbox{m}^2$ atomic force microscope
image of the Au coating on the plate. The topography of the
coating on the sphere is similar.}

\end{figure*}
\begin{figure*}
\vspace*{-5cm}
\includegraphics{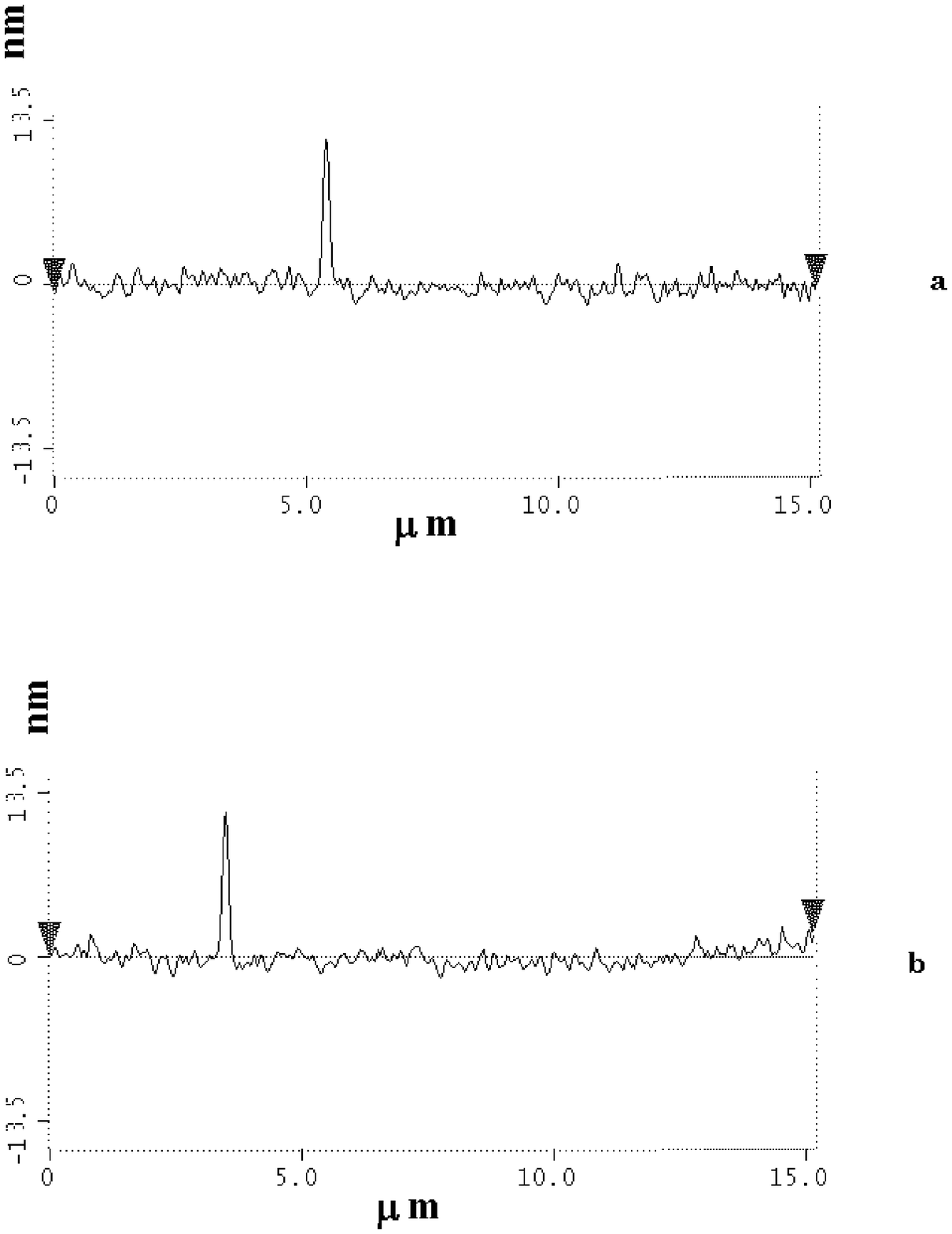}
\vspace*{-5cm}
\caption{Typical cross sections of the atomic force microscope
image of the Au coating on the plate with (a) constant $x$  
and (b) constant $y$.
}
\end{figure*}
\end{document}